\documentstyle[proceedings]{crckapb}
\begin{opening}
\title{The Virgo Consortium: The evolution and formation of galaxy clusters}
\author{J.\ M.\ Colberg}
\author{S.\ D.\ M.\ White}
\institute{Max--Planck--Institut f\"ur Astrophysik, Karl--Schwarzschild--Str.\ 1,
85740 Garching, Germany}
\author{A.\ Jenkins}
\author{F.\ R.\ Pearce}
\author{C.\ S.\ Frenk}
\institute{Physics Dept., University of Durham, Durham DH1 3LE, UK}
\author{P.\ A.\ Thomas}
\author{R.\ Hutchings}
\institute{MAPS, University of Sussex, Brighton BN1 9QH, UK}
\author{H.\ M.\ P.\ Couchman}
\institute{Dept.\ of Astronomy, University of Western Ontario, London, 
Ontario N6A 3K7, Canada}
\author{J.\ A.\ Peacock}
\institute{Royal Observatory, Blackford Hill, Edinburgh EH9 3HJ, UK}
\author{G.\ P.\ Efstathiou}
\institute{Dept.\ of Physics, Nuclear Physics Building, Keble Road,
Oxford OX1 3RH, UK}
\author{A.\ H.\ Nelson}
\institute{Dept.\ of Physics and Astronomy, UWCC, PO Box 913,
Cardiff, UK}
\end{opening}
\begin{document}
\begin{abstract}
We report on work done by the Virgo consortium, an international
collaboration set up in order to study the formation and evolution
of Large Scale Structure using N--body simulations on the latest generation
of parallel supercomputers. We show results of $256^3$ particle simulations
of the formation of clusters in four Dark Matter models with different
cosmological parameters. Normalizing the models such that one obtains the
correct abundance of rich clusters yields an interesting result: The
peculiar velocities of the clusters are almost independent of
$\Omega$, and depend only weakly
on $\Gamma$, the shape parameter of the power spectrum. Thus, it is 
nearly impossible to distinguish between high and low $\Omega$ models
on the basis of the peculiar velocities.
\end{abstract}
\section{Introduction} \label{intro}
Clusters of galaxies are the largest objects one finds today in the
Universe. They are well studied from both observational
and theoretical viewpoints (for a recent review
c.f.\ \citeauthor{Bahcall96a} (\citeyear{Bahcall96a})). 

It may be possibile to distinguish between high and low
$\Omega$ Universes by using the peculiar velocities
of the clusters. On this basis \citeauthor{Bahcall96} 
(\citeyear{Bahcall96}) concluded that
the observed clusters velocity function ''is most consistent
with a low mass-density ($\Omega\sim 0.3$) CDM model''. However, we
show in section \ref{pvlt} that this is dubious when 
models are normalized such that they all give the correct {\em abundance\/}
of rich clusters. We present results from the
simulations that support our theoretical considerations
in section \ref{pvs}.

Before we talk about the
details of the analysis of clusters from the simulations we give in
section \ref{code} a brief description of the numerical code. In section
\ref{simulations} we discuss the models used.

\section{The Code} \label{code}

The Virgo consortium was formed in order to study the evolution of
both Dark Matter and gas in the expanding Universe using the latest
generation of parallel supercomputers.

The code used is called ''Hydra'' which was developed by 
\citeauthor{Couchman95} (\citeyear{Couchman95})
and parallelized by \citeauthor{Pearce95} 
(\citeyear{Pearce95}). Gravity is treated using an
adaptive P$^3$M technique. This places mesh refinements on the
regions of strongest clustering. Large refinements are done in
parallel across the processors, smaller ones are farmed out to
single processors. 

The simulations were run on the Cray T3D supercomputers at the
computer center of the Max Planck Society in Garching and at
the EPCC in Edinburgh.

\section{The Dark Matter Simulations} \label{simulations}

We have carried out a set of very large N--body simulations of CDM
universes with four different choices of parameters. The models
chosen were Standard CDM (SCDM), a high $\Omega$ model with an
additional radiative component ($\tau$CDM), an Open CDM model 
(OCDM) and a flat low $\Omega$ model with a Cosmological
Constant ($\Lambda$CDM). With the 
exception of the Open Model, which only had 200$^3$ particles, each simulation 
followed the evolution of 256$^3$ particles in a box of
239.5\,$h^{-1}$\,Mpc\footnote{As usual we express the Hubble
constant as $H_0 = 100\,h\,$km/(Mpc\,sec).} on a side. 

\begin{table}
\caption{The Virgo models}
\begin{center}
\begin{tabular}{lccccc}
\hline
Model & $\Omega$ & $\Lambda$ & $h$ & $\sigma_8$ & $\Gamma$ \\
\hline
SCDM & 1.0 & 0.0 & 0.5 & 0.51 & 0.50 \\
$\tau$CDM & 1.0 & 0.0 & 0.5 & 0.51 & 0.21 \\
$\Lambda$CDM & 0.3 & 0.7 & 0.7 & 0.90 & 0.21 \\
OCDM & 0.3 & 0.0 & 0.7 & 0.85 & 0.21 \\
\hline
\end{tabular}
\end{center}
\end{table}

In all models, the initial fluctuation amplitude was set by
requiring that the models should reproduce the observed
abundance of rich clusters (for details see section \ref{pvlt}).
Table 1 gives the parameters of the models.

\begin{figure}
\vspace{170mm}
\label{clusterpic}
\caption{The evolution of the same cluster in the $\tau$CDM (leftmost
pictures) and in the $\Lambda$CDM model (rightmost pictures), for
redshifts of 2 (top), 1 (middle), and 0 (bottom). The sizes of the
regions shown are $21\times 21\times 8\,h^{-1}$\,Mpc$^3$ and
$35\times 35\times 13\,h^{-1}$\,Mpc$^3$ for $\tau$CDM and $\Lambda$CDM,
respectively.}
\end{figure}

Figure \ref{clusterpic} shows the evolution of the same 
cluster\footnote{The simulations were run with the same phases.}
in the $\tau$CDM and the $\Lambda$CDM model for the redshifts
2, 1, and 0. Structure forms earlier in the low $\Omega$ universe.
But already at a redshift of 2 a very large filamentary structure
can be seen in the high $\Omega$ universe, 
too\footnote{The clusters were taken from an additional set of runs
with the same parameters as above but box sizes of 85\,$h^{-1}$\,Mpc
and 141\,$h^{-1}$\,Mpc for the high and low $\Omega$ models, respectively.}.

\section{Peculiar Velocities of Galaxy Clusters from Linear Theory}
\label{pvlt}

According to linear theory, the mean--square peculiar velocity of
galaxy clusters is
\begin{equation}
\langle v^2_{3D}\rangle = \frac{1}{2\pi^2}\, H_0^2\, \Omega^{1.2} \int_{0}^{\infty} 
P(k,\Gamma)\, W^2(k,R)\, dk\,,
\label{pecvol}
\end{equation}
where $W^2(k,R)=\exp[-k^2\,R^2]$ is a window function of radius $R$ (see,
e.g., \citeauthor{Croft94} 
(\citeyear{Croft94})). The power spectrum $P(k)$  is taken in its
parametric form introduced by \citeauthor{Bond84} 
(\citeyear{Bond84}),
\begin{equation}
P(k) = \frac{B k}{\{1+[ak+(bk)^{3/2}+(ck)^2]^{\nu}\}^{2/\nu}}\,,
\label{power}
\end{equation}
where $a=(6.4/\Gamma)\,h^{-1}\,$Mpc, $b=(3.0/\Gamma)\,h^{-1}\,$Mpc, $c=(1.7/\Gamma)\,h^{-1}\,$Mpc 
and $\nu=1.13$. The quantity $\Gamma$ is given by
\begin{equation}
\Gamma = \left\{ \begin{array}{ll}
                 \Omega_0 h & \mbox{models with $\Omega_0 + \Lambda_0$} = 1 \\
                 \Omega_0 h/[0.861+3.8(m_{10}\tau_{d})^{2/3}]^{1/2} & 
                        \mbox{models with decaying $\nu$} 
                 \end{array}
         \right.
\label{gamma}
\end{equation}
where $m_{10}$ is the neutrino mass in units of 10\,keV and $\tau_{d}$ is its lifetime
in years (\citeauthor{Bond91} (\citeyear{Bond91})). 
Equation (\ref{gamma}) is valid for any model in a spatially flat universe.

The normalisation $B$ of the power spectrum can be obtained in two
different ways: Either by
using a relationship between $B$ and the COBE measurements (as shown
in \citeauthor{Efstathiou92} (\citeyear{Efstathiou92})) or by
relating $B$ to the rms linear fluctuation in the mass distribution on scales of
$8\,h^{-1}$\,Mpc, $\sigma_8$, which is defined by
\begin{equation}
\sigma_8^2 \equiv \frac{1}{(2\pi)^3} \int_{0}^{\infty} P(k,\Gamma) 
\left( \frac{3}{k R_8} j_1(k R_8)\right)^2 d^3k\,,
\label{sigma8}
\end{equation}
where $R_8 \equiv 8\,h^{-1}\,$Mpc, and $j_1$ is a spherical Bessel function.

Values for $\sigma_8$ were obtained by either using
the mass or the X--ray temperature functions 
of rich clusters (\citeauthor{White93} 
(\citeyear{White93}) (WEF), \citeauthor{Viana95} (\citeyear{Viana95}),
\citeauthor{Eke96} (\citeyear{Eke96}) (ECF)). 
WEF obtain $\sigma_8 \approx 0.57\, \Omega_0^{-0.56}$,
ECF get
\begin{equation}
\sigma_8 = \left\{ \begin{array}{ll}
                   (0.52\pm 0.04)\,\Omega_0^{-0.46+0.10\,\Omega_0} & 
                        \mbox{for $\Lambda_0=0$} \\
                   (0.52\pm 0.04)\,\Omega_0^{-0.52+0.13\,\Omega_0} &
                        \mbox{for $\Omega_0+\Lambda_0=1$}
                  \end{array}
             \right.
\label{sigma8ecf}
\end{equation}
with the quoted statistical uncertainties obtained using bootstrap
methods. 
 
Inserting eq.\ (\ref{sigma8}) into eq.\ (\ref{pecvol}) yields
\begin{equation}
\langle v^2_{3D}\rangle = 10^4\, (\Omega^{0.6}\, \sigma_8)^2\,
f(\Gamma,R)\quad [km/sec]\,,
\label{pecvol2}
\end{equation}
with the abbreviation
\begin{equation}
f(\Gamma,R)\equiv 4\pi h^2
\frac{\int P(k) W^2(k,R)dk}{\int P(k)\left( \frac{3}{k R_8} j_1(k R_8)\right)^2 d^3k}
\end{equation}
Comparing eqs.\ (\ref{sigma8ecf}) and (\ref{pecvol2}) one sees that 
$(\Omega^{0.6} \sigma_8)^2$ only very weakly depends on $\Omega$.
Thus, the peculiar velocities of galaxy clusters are nearly a function of
$\Gamma$ alone.

Taking $R=1.5\,h^{-1}$\,Mpc one obtains
\begin{equation}
f(\Gamma,R=1.5\,h^{-1}\,\mbox{Mpc}) = 12.5\,\Gamma^{-1.08} +
49.4
\label{fgamma}
\end{equation}

\section{Peculiar Velocities of Galaxy Clusters from the Simulations}
\label{pvs}

When comparing
N--body simulations with real data one has to find a way to
select the clusters in the simulation. WEF find clusters
by locating high--density regions with a friends--of--friends group
finder, and then getting masses from the particle count within spheres
of comoving radius $r=1.5\,h^{-1}$\,Mpc.
We use the same algorithm and treat objects with a mass larger
than $5.5\cdot 10^{14} h^{-1} \mbox{M}_{\odot}$ as galaxy
clusters.

\begin{table}
\begin{center}
\caption{Peculiar velocities of galaxy clusters}
\begin{tabular}{lccc}
\hline
Model & $\langle v^2_{3D}\rangle_{lt}$ [km/sec] & $\langle v^2_{3D}\rangle_{Sim}$
[km/sec] & $N$ \\
\hline
SCDM & 441 & 461 & 20 \\
$\tau$CDM & 540 & 492 & 21 \\
$\Lambda$CDM & 462 & 452 & 21  \\
OCDM & 437 & 476 & 18 \\
\hline
\end{tabular}
\end{center}
\end{table}

In table 2 we give the peculiar velocities for the four
models from linear theory (first column, using eq.s\ (\ref{pecvol2})
and (\ref{fgamma})), from the simulations (second column), and the
number of clusters found in the simulations (third column).
Linear theory clearly predicts the simulation results to within
their uncertainties.

From the numbers it is obvious that the peculiar velocities are not
a good way to discriminate between the models if these are normalized
such that they reproduce the correct abundance of rich clusters. In
particular, there is no difference between high and low
$\Omega$ models.

\section*{Acknowledgements}

JMC would like to thank Matthias Bartelmann and Antonaldo
Diaferio for numerous helpful and interesting discussions.

\end{document}